\documentclass[superscriptaddress,twocolumn,showpacs,amsmath,amssymb]{revtex4}
\usepackage{graphicx}
\usepackage{color}

\renewcommand{\vec}[1]{\mbox{\boldmath $#1$}}

\begin{document}

\title{
Semi-microscopic modeling of heavy-ion fusion reactions \\
with multi-reference covariant density functional theory
}

\author{K. Hagino}
\affiliation{
Department of Physics, Tohoku University, Sendai 980-8578,  Japan}
\affiliation{Research Center for Electron Photon Science, Tohoku
University, 1-2-1 Mikamine, Sendai 982-0826, Japan}
\affiliation{
National Astronomical Observatory of Japan, 2-21-1 Osawa,
Mitaka, Tokyo 181-8588, Japan}

\author{J. M. Yao}
\affiliation{
Department of Physics, Tohoku University, Sendai 980-8578,  Japan}
\affiliation{School of Physical Science and Technology,
             Southwest University, Chongqing 400715, China}


\begin{abstract}
We describe low-lying collective excitations of atomic nuclei with
the multi-reference covariant density functional theory,
and combine them with coupled-channels calculations for heavy-ion fusion
reactions at energies around the Coulomb barrier.
To this end, we use the calculated transition
strengths among several collective states as inputs to the coupled-channels
calculations.
This approach provides a natural way to describe anharmonic multi-phonon
excitations as well as a deviation of rotational excitations
from a simple rigid rotor.
We apply this method to subbarrier fusion reactions of $^{58}$Ni+$^{58}$Ni,
$^{58}$Ni+$^{60}$Ni and $^{40}$Ca+$^{58}$Ni systems.
We find that the effect of anharmonicity tends to smear the fusion barrier
distributions, better reproducing the experimental data
compared to the calculations in the harmonic oscillator limit.
\end{abstract}


\maketitle

\section{Introduction}

In heavy-ion fusion reactions at energies around the Coulomb barrier,
low-lying collective excitations of the colliding nuclei
during fusion process plays an important role
in enhancing fusion cross sections as compared to a prediction of a simple potential model
\cite{DHRS98,BT98,Back14}.
These effects have usually been taken into account in the
coupled-channels calculations with a coupling scheme
based on a harmonic vibrator for spherical nuclei or on a
rigid rotor for deformed nuclei \cite{HRK99,HT12}.
With this approach, the energy of the first excited state
as well as the coupling strength from the ground state to the
first excited state, which can be taken from the experimental data,
specify all the other excitation energies and coupling strengths for higher members in
the coupling scheme.
Typical examples which show the subbarrier enhancement 
of fusion cross sections include the fusion of
$^{58}$Ni+$^{60}$Ni and $^{64}$Ni+$^{64}$Ni, for which multi-phonon
excitations have been shown to play an important
role \cite{Stefanini95,Esbensen05}. See also
Refs. \cite{TI86,KRNR93} for discussions on
multi-phonon excitations.
Multiple excitations within the ground state rotational band
also play an important role in most of fusion reactions involved with
heavy deformed nuclei \cite{LDH95}.

In reality, however, most of atomic nuclei have neither a pure harmonic oscillator
spectrum nor a pure rigid body rotational band.
For example, the $^{58}$Ni nucleus, which has usually been
considered to be a typical vibrational nucleus,
does not exhibit a level spectrum characteristic to the
harmonic vibration, that is, the degeneracy of the two-phonon
triplet is considerably broken.
Moreover, a recent theoretical 
calculation based on a multi-reference density
functional theory
with the Skyrme interaction also indicates that the $B(E2)$ strengths
among the collective levels in $^{58}$Ni 
deviate largely from what are expected
from a simple harmonic oscillator \cite{Yao15}. 
It is therefore of considerable interest to investigate
the role of anharmonicity, that is, the deviation from the
harmonic limit, in subbarrier fusion of $^{58}$Ni.

In Refs. \cite{HTK97,HKT98}, the effect of anharmonicity on
subbarrier fusion of $^{16}$O+$^{144,148}$Sm has been investigated using
the vibrational limit of interacting boson model (IBM). See also
Ref. \cite{Zamrun08} for an application of this method to
large-angle quasi-elastic scattering of the $^{16}$O+$^{144}$Sm system.
Although the static quadrupole moment of the first
excited state could be successfully
extracted by analyzing the high precision experimental
data with this approach \cite{HTK97,HKT98},
the application of IBM has several limitations for a global study.
Firstly, the method is not applicable to doubly magic nuclei, as the number
of bosons in IBM is estimated from the number of valence nucleons outside
shell closures. Secondly, phenomenological parameters have to be
introduced to the model Hamiltonian and to the transition operators.
It is therefore desirable to develop an alternative microscopic
approach for nuclear collective excitations, which does not rely on
the harmonic limit or the rigid rotor, in order to systematically investigate
the effect of collective excitations in general on subbarrier fusion reactions
in a wide mass region.

In this paper, we employ a beyond-mean-field method to describe
low-lying collective excitations and combine it with the coupled-channels approach to heavy-ion fusion reactions.
The pure mean-field approximation breaks the rotational symmetry
and does not yield a spectrum of nuclei.
This can be overcome by going beyond the mean-field approximation,
in particular, by carrying out the angular momentum projection.
One can also take into account the quantum fluctuation of the mean-field
wave function by superposing many Slater determinants with the
generator coordinate method (GCM). When the pairing correlation is
important, the particle number projection can also be implemented.
Such scheme has been referred to as a multi-reference density-functional
theory (MR-DFT),
and has rapidly been developed in the nuclear structure physics
for the past decade \cite{Bender03,Yao14}.

The paper is organized as follows.
In Sec. II, we briefly review the coupled-channels approach for
heavy-ion sub-barrier fusion reactions.
In Sec. III, we present the results of MR-DFT calculations for the
$^{58}$Ni and $^{60}$Ni nuclei. To this end, we use the covariant
density functional theory (CDFT), based on the relativistic
framework.
In Sec. IV, we combine the coupled-channels calculations with
the MR-CDFT approach. We apply this method to subbarrier fusion
reactions of $^{58}$Ni+$^{58}$Ni, $^{58}$Ni+$^{60}$Ni, and
$^{40}$Ca+$^{58}$Ni systems, and discuss the role of anharmonicity
of quadrupole vibrations of $^{58}$Ni and $^{60}$Ni.
We then summarize the paper in Sec. V.

\section{Coupled-channels approach to heavy-ion fusion reactions}

Our aim in this paper is to solve coupled-channels equations
using inputs from the MR-CDFT calculations.
In principle, one could formulate the coupled-channels method
fully microscopically using the MR-CDFT method.
In such approach, the internuclear potentials, both for the diagonal
and the coupling parts, would be constructed by folding an effective
nucleon-nucleon interaction with calculated density distributions and
transition densities \cite{FHTYS12}. 
It has been known, however, that this double folding
procedure fails to work for heavy-ion subbarrier
fusion reactions \cite{HT12}. That is,
one obtains a surface diffuseness parameter of around $a\sim$ 0.63 fm
when a double folding potential is
fitted with a Woods-Saxon function, whereas experimental fusion
cross sections systematically require a much larger value, {\it e.g.,}
$a\sim$ 1.0 fm \cite{LDH95,Newton04,Gontcher04,HRD03,Ei12,MHD07}.
An important fact to notice is that
the double folding method works only in the surface region of the
potential.
For elastic and inelastic
scattering, 
the surface region of the potential is mainly probed and a double
folding potential is reasonable \cite{MHD07,HTBT05,WHD06,GEH07,EDH08}. 
In marked contrast,
fusion reactions involve both the surface and the inner regions, where
two nuclei appreciably overlap with each other.
As a consequence, several dynamical effects are
important in the inner region \cite{IHI07-systematic,HW07,IHI07,IM13,ME06}, and the double folding
potential looses its validity.

Another problem of the fully microscopic formulation is that the MR-CDFT
calculations seldom yield a perfect agreement with experimental data
for excitation energies and transition strengths, even though an overall
agreement is often reasonable.
On the other hand, in order to describe quantitatively heavy-ion fusion
reactions at energies close to the Coulomb barrier, it is important to use
reasonable values for excitation energies and transition strengths.

In order to avoid these drawbacks of the fully microscopic approach,
in this paper we employ a semi-microscopic approach. That is,
we use a phenomenological Woods-Saxon internuclear potential and adopt the
experimental value for the coupling strength between the ground state
and the first excited state. The coupling strengths for higher members
are not known well in many nuclei, and it is for these values that
we employ the MR-CDFT calculations, after scaling the calculated values
with the experimental strength for the transition between the ground
state and the first excited state. The excitation energies are known for
most of the collective levels, and we simply use the experimental values
whenever they are available.

In the coupled-channels approach to subbarrier fusion reactions,
one expands the total wave function of the system in terms of the eigen-functions
of the collective states in the target nucleus, $|\varphi_{J0}\rangle$,
as
\begin{equation}
  \Psi_{LM_L}(\vec{r})=\sum_J\frac{u_J(r)}{r}Y_{LM_L}(\hat{\vec{r}})
  |\varphi_{J0}\rangle,
\label{totwf}
\end{equation}
where $r$ is the relative coordinate between the colliding nuclei, and
$J$ and $L$ are the angular momentum for the target state and the
angular momentum for the relative motion, respectively.
Here, for simplicity of notation,
we have assumed that the projectile nucleus is inert, but an extension
is straightforward to the case where both the projectile and the target nuclei
are excited. We have also introduced the isocentrifugal
approximation \cite{HT12}, and have assumed that the angular momentum for
the relative motion does not change by
the excitation of the target nucleus. Notice that only the $J_z=0$ component is
excited in the target nucleus in the isocentrifugal approximation.

Substituting Eq. (\ref{totwf}) to the projected Schr\"odinger
equation for the energy $E$, that is,
$\langle \varphi_{J0}|H-E|\Psi_{LM_L}\rangle=0$, where $H$ is the total Hamiltonian,
one obtains the coupled-channels equations for the radial wave functions
$u_J(r)$ as\cite{HT12},
\begin{eqnarray}
&&\left[-\frac{\hbar^2}{2\mu}\frac{d^2}{dr^2}
    +\frac{L(L+1)\hbar^2}{2\mu r^2}
    +V_0(r)
-E+\epsilon_J\right]u_J(r) \nonumber \\
&&+\sum_{J'}V_{JJ'}(r)u_{J'}(r)=0,
\label{cceq}
\end{eqnarray}
where $\mu$ is the reduced mass for the relative motion, $V_0(r)$ is the
bare potential, and $\epsilon_J$ is the energy of the target state $J$.
$V_{JJ'}(r)$ are the coupling matrix elements given by
\begin{equation}
  V_{JJ'}(r)=\langle\varphi_{J0}|V_{\rm coup}(r,\alpha_{\lambda0})|\varphi_{J'0}
  \rangle,
\label{coupV}
\end{equation}
where $V_{\rm coup}$ is the coupling potential and $\alpha_{\lambda 0}$ is
the excitation operator with a multipolarity $\lambda$.
We solve the coupled-channels equations
by imposing the incoming wave
boundary condition at $r=r_{\rm abs}$ inside the Coulomb barrier
\cite{HT12}, that is,
\begin{eqnarray}
u_J(r)&\sim&
\sqrt{\frac{k_{J_0}}{k_J(r)}}\,{\cal T}^L_{JJ_0}
\exp\left(-i\int^r_{r_{\rm abs}}k_J(r')dr'\right)
~~(r \leq r_{\rm abs}), \nonumber \\
\\
&=& H_L^{(-)}(k_Jr)\delta_{J,J_0}
-\sqrt{\frac{k_{J_0}}{k_J}}\,{\cal S}^J_{JJ_0}H_L^{(+)}(k_Jr)
~~(r\to\infty), \nonumber \\
\end{eqnarray}
where $H_L^{(+)}$ and $H_L^{(-)}$ are the outgoing and the incoming
Coulomb wave functions, respectively.
${\cal S}^L_{JJ_0}$ and ${\cal T}^L_{JJ_0}$ are the nuclear $S$-matrix and the
transmission coefficient, respectively, with $J_0=0$ being the spin of the
ground state of the target nucleus.
$k_J(r)$ is the local wave number given by
\begin{equation}
k_J(r)=\sqrt{
\frac{2\mu}{\hbar^2}\left(E-\epsilon_J-\frac{L(L+1)\hbar^2}{2\mu r^2}
-V_0(r)\right)},
\end{equation}
whereas $k_J=k_J(r=\infty)=\sqrt{2\mu(E-\epsilon_J)/\hbar^2}$.
The fusion cross section $\sigma_{\rm fus}$ is then obtained as
\begin{equation}
  \sigma_{\rm fus}(E)=\frac{\pi}{k^2}\sum_L(2L+1)P_L(E),
\end{equation}
with $P_L(E)=\sum_J \vert {\cal T}^L_{JJ_0} \vert ^2$.

As we have mentioned, we employ the Woods-Saxon potential for the
nuclear part of the bare potential, $V^{(N)}_0(r)$, that is,
\begin{equation}
  V^{(N)}_0(r)=-\frac{V_0}{1+\exp[(r-R_0)/a]}.
\end{equation}
The nuclear coupling potential is obtained by
deforming the radius $R_0$ to
\begin{equation}
  R_0\to R_0+R_T\sum_\mu\alpha_{\lambda\mu}Y_{\lambda\mu}^*(\hat{\vec{r}}),
\end{equation}
where $R_T$ is the radius of the target nucleus. Here, the deformation
parameter $\alpha_{\lambda\mu}$ is related to the electric multipole
operator as \cite{Ring1980},
\begin{equation}
  Q_{\lambda\mu}=\frac{3e}{4\pi}Z_TR_T^2\alpha_{\lambda\mu},
\label{q}
\end{equation}
where $Z_T$ is the atomic number of the target nucleus.
The coupling potential in Eq. (\ref{coupV})
in the isocentrifugal approximation then reads \cite{HT12},
\begin{eqnarray}
V_{\rm coup}(r,\alpha_{\lambda0})&=&
-\frac{V_0}{1+\exp[(r-R_0-\sqrt{\frac{2\lambda+1}{4\pi}}
    R_T\alpha_{\lambda0})/a]} \nonumber \\
&&+\frac{3}{2\lambda+1}Z_PZ_Te^2\frac{R_T^\lambda}{r^{\lambda+1}}
\sqrt{\frac{2\lambda+1}{4\pi}}\alpha_{\lambda0},\nonumber \\
&&-V_0^{(N)}(r),
\end{eqnarray}
where we have also included the Coulomb coupling potential.
The last term is to avoid the double counting in Eq. (\ref{cceq}).

The matrix elements of $V_{\rm coup}$, Eq. (\ref{coupV}), can be
evaluated with the method employed in the computer 
code {\tt CCFULL} \cite{HRK99,HT12}.
To this end, one needs the matrix elements of the operator
$\sqrt{\frac{2\lambda+1}{4\pi}}\alpha_{\lambda0}$.
In the following, we define the coupling strengths
for the coupled-channels calculations
as
\begin{equation}
  \frac{\beta_{JJ'}^{(\lambda)}}{\sqrt{4\pi}}\equiv
  \sqrt{\frac{2\lambda+1}{4\pi}}\,\langle\varphi_{J0}|\alpha_{\lambda0}|
  \varphi_{J'0}\rangle.
\label{beta}
\end{equation}
For the quadrupole harmonic oscillator with $\lambda=2$,
this definition yields \cite{HT12}
\begin{equation}
  \beta^{(\lambda=2)}_{J2_1}=\sqrt{2}\,\beta^{(\lambda=2)}_{2_10_1}\langle 2020|J0\rangle,
\end{equation}
for the coupling between the one phonon state (that is, 
the 2$^+_1$ state) to the
two phonon state with the angular momentum $J$.
Notice that $\sqrt{\sum_{J=0,2,4}(\beta^{(\lambda=2)}_{J2_1})^2}=\sqrt{2}\,\beta^{(\lambda=2)}
_{2_10_1}$, which has often
been employed in the coupled-channels calculations with
multi-phonon couplings \cite{HRK99,HT12,Esbensen05,TI86,KRNR93}.
For a rigid rotor, one obtains
\begin{equation}
\beta^{(\lambda=2)}_{22}=\frac{2\sqrt{5}}{7}\beta,
  ~~\beta^{(\lambda=2)}_{24}=\frac{6}{7}\beta,
  ~~\beta^{(\lambda=2)}_{44}=\frac{20\sqrt{5}}{77}\beta,
\end{equation}
with $\beta^{(\lambda=2)}_{20}\equiv\beta$ (see Eq. (3.49) in Ref. \cite{HT12}).

With microscopic nuclear structure calculations, the coupling strengths
can be estimated as (see Eqs. (\ref{q}) and (\ref{beta})),
\begin{equation}
  \frac{\beta_{JJ'}^{(\lambda)}}{\sqrt{4\pi}}=
  \sqrt{\frac{2\lambda+1}{4\pi}}\,
\frac{4\pi}{3Z_TeR_T^2}
  \langle\varphi_{J0}|Q_{\lambda0}|\varphi_{J'0}\rangle,
  \label{beta-gcm}
\end{equation}
where
the quantity $\langle\varphi_{J0}|Q_{\lambda0}|\varphi_{J'0}\rangle$ can be
evaluated microscopically using the operator
$Q_{\lambda\mu}=\sum_ir_i^\lambda Y_{\lambda\mu}(\hat{\vec{r}}_i)$.

In coupled-channels calculations for heavy-ion reactions, one sometimes
uses a different value of the nuclear coupling
strength from the Coulomb coupling strength, 
see {\it e.g.,} Ref. \cite{Esbensen05}.
While the Coulomb coupling strength, $\beta_C$,
can be estimated from a measured electric
transition strengths, $B(E\lambda$),
the nuclear coupling strengths, $\beta_N$, are taken rather arbitrary.
One of the big advantages of the semi-microscopic approach is that
the nuclear coupling strengths can also be estimated by using the isoscalar
operator for $Q_{\lambda\mu}$, that is,
\begin{equation}
  Q^{\rm (IS)}_{\lambda\mu}=\sum_{i\in p,n}r_i^\lambda Y_{\lambda\mu}(\hat{\vec{r}}_i)
  =\frac{3}{4\pi}A_TR_T^2\alpha_{\lambda\mu},
\end{equation}
whereas the Coulomb coupling strengths are related to the $E\lambda$ operator,
\begin{equation}
  Q^{\rm (E\lambda)}_{\lambda\mu}=e\sum_{i\in p}r_i^\lambda Y_{\lambda\mu}(\hat{\vec{r}}_i)
  =\frac{3e}{4\pi}Z_TR_T^2\alpha_{\lambda\mu}.
\end{equation}
From these equations, one obtains
\begin{equation}
  \frac{\beta_N}{\beta_C}=\frac{Z_T}{A_T}\left(1+\frac{M_n}{M_p}\right),
  \label{mn/mp}
\end{equation}
where
$M_n/M_p$ is the neutron-to-proton ratio for a transition given by
\begin{equation}
\frac{M_n}{M_p}=\frac
{\langle \varphi_{JM}|\sum_{i\in n}r_i^\lambda Y_{\lambda\mu}(\hat{\vec{r}}_i)|\varphi_{J'M'}\rangle}
{\langle \varphi_{JM}|\sum_{i\in p}r_i^\lambda Y_{\lambda\mu}(\hat{\vec{r}}_i)|\varphi_{J'M'}\rangle}.
\end{equation}
Notice that for a pure isoscalar transition,
$M_n/M_p$ is reduced to $N_T/Z_T$, and the nuclear and the Coulomb
coupling strengths are identical to each other, that is,
$\beta_N=\beta_C$.

\section{Multi-reference covariant density functional calculation
  for $^{58}$N\lowercase{i} and $^{60}$N\lowercase{i}}

Let us now carry out the MR-CDFT calculations for the 
$^{58}$Ni and $^{60}$Ni nuclei and obtain inputs for the coupled-channels
calculations. These nuclei have been studied recently in Ref. \cite{Yao15}
using the non-relativistic MR-DFT method. Here we repeat similar calculations
with the relativistic framework, in order to check
the dependence of the conclusions on a choice of energy density functional. 

In the MR-CDFT~\cite{Yao14,MRCDFT1,MRCDFT2,MRCDFT3},
the wave function for nuclear low-lying states is constructed as a superposition
of projected mean-field states corresponding to different
deformations parameter $\beta$,
\begin{equation}
 \label{TrialWF}
 \vert \alpha JM; NZ\rangle
 =\sum_{\beta} f^{J}_\alpha(\beta) \hat P^J_{M0} \hat P^N\hat P^Z\vert \Phi(\beta) \rangle,
\end{equation}
where $\alpha=1,2,\ldots$ distinguishes different collective
states with the same angular momentum $J$.
Here, $\vert \Phi(\beta) \rangle$ are the mean-field states 
generated by
the deformation constrained relativistic mean-field (RMF) method.
The pairing correlation is taken into account in the BCS approximation.
For simplicity, we have assumed the axial and reflection symmetries
for the mean-field states.
Thus, the $K$ quantum number (that is, the
projection of angular momentum onto the $z$-axis) is zero, and
the index $K$ has been dropped. 
We implement both the particle-number and the angular-momentum
projections.
The projection operators $\hat{P}^{N}$, $\hat{P}^{Z}$, and $\hat P^J$ in
Eq. (\ref{TrialWF}) 
project
the mean-field states 
onto the states with good neutron and proton numbers $N, Z$ as well as a
good angular momentum $J$.

The weight coefficients $f^{J}_\alpha(\beta)$ in Eq. (\ref{TrialWF})
are determined by
solving the Hill-Wheeler-Griffin (HWG) equations~\cite{Ring1980},
\begin{eqnarray}
\label{HWGeq}
  \sum_{\beta'}\left[ {\cal H}^{J}_{00}(\beta,\beta')-E^J_\alpha {\cal N}^{J}_{00}(\beta,\beta')\right] f^{J}_\alpha(\beta')=0,
\end{eqnarray}
where  ${\cal N}^{J}_{00}(\beta,\beta')=\langle \Phi(\beta)\vert\hat P^J_{00} \hat P^N\hat P^Z\vert \Phi(\beta')\rangle$
and ${\cal H}^{J}_{00}(\beta,\beta')=\langle \Phi(\beta)\vert\hat H\hat P^J_{00} \hat P^N\hat P^Z\vert \Phi(\beta')\rangle$
are the norm and the energy kernels, respectively. The prescription
of mixed density is adopted for the energy kernel.
The detailed expressions for the kernels can be found in Refs.
~\cite{Yao14,MRCDFT1,MRCDFT2,MRCDFT3}.
The solution of the HWG equation provides the energy levels
and the information on the matrix elements
$\langle\varphi_{J0}|Q_{\lambda0}|\varphi_{J'0}\rangle$ in Eq.(\ref{beta-gcm})
with $|\varphi_{JM}\rangle=|\alpha JM;NZ\rangle$, which are  
needed for the coupled-channels calculations for fusion cross sections.
Notice that in most of cases it is hard to determine experimentally the
sign of this matrix element even if information is sometimes available
on its absolute value from the electric
transition probabilities \cite{Esbensen05} (see also Ref. \cite{Zamrun10}). 
An advantage of the present semi-microscopic approach is that the matrix
elements can be estimated theoretically, including their sign as well. 

\begin{figure}
\includegraphics[width=8.6cm]{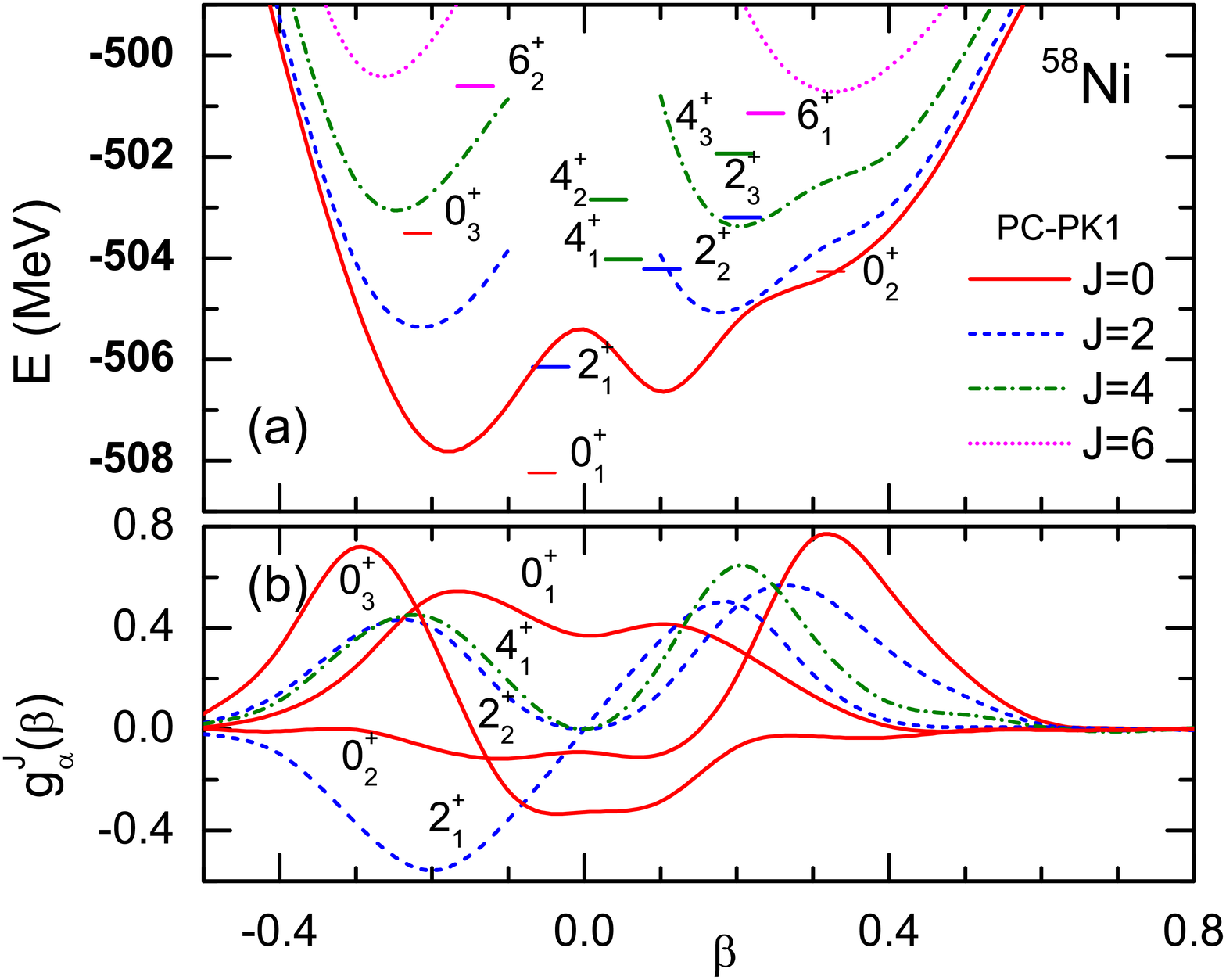}
\caption{(Color online) (a) The total energy for
  the angular-momentum-projected
  states with
  $J = 0, 2, 4$ and 6
  for $^{58}$Ni as a function of the intrinsic mass quadrupole
  deformation $\beta$ of the mean-field states.
  The particle number projection has also been implemented.
  Those curves are obtained with the projected
  CDFT method with PC-PK1 interaction~\cite{Zhao10}.
  The low-lying collective
  levels, obtained with the configuration mixing calculation, 
  are also plotted at their average deformation $\bar\beta$.
  (b) The collective wave functions given by Eq. (\ref{collwf})
  for the states indicated in the figure as a function of deformation
  parameter $\beta$. }
\label{Ni58PES}
\end{figure}

Figure~\ref{Ni58PES} (a)
shows the energy curves of
particle-number and angular-momentum-projected states
with $J = 0, 2, 4$ and 6 for $^{58}$Ni as a function of
the intrinsic mass quadrupole deformation $\beta$ of the mean-field
states.
The PC-PK1 parameter set \cite{Zhao10} is used for the
nucleon-nucleon interaction. 
The figure also shows the 
low-lying collective states obtained by mixing the symmetry
conserved states.
These states are plotted at their average deformation $\bar{\beta}$
defined as
\begin{equation}
  \bar{\beta}(J_\alpha)=\sum_\beta \beta |g^J_\alpha(\beta)|^2,
\end{equation}
where 
the collective wave functions $g^J_\alpha(\beta)$ are related to the
weight functions in Eq.(\ref{TrialWF}) as
\begin{equation}
  g^{J}_{\alpha}(\beta)= \sum_{\beta'} [{\cal N}^{J}_{00}]^{1/2}(\beta,\beta')
  f^{J}_{\alpha} (\beta').
\label{collwf}
\end{equation}
These collective wave functions are plotted in the lower panel of
Fig. ~\ref{Ni58PES}.

\begin{figure}
  \includegraphics[width=8cm]{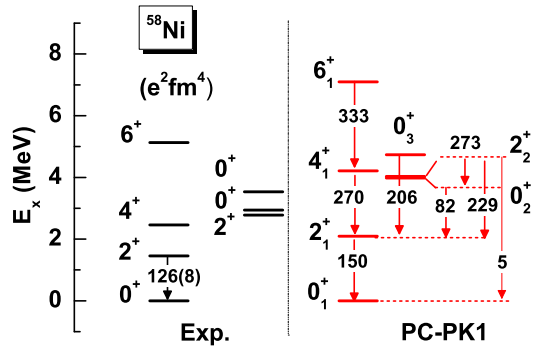}
\caption{(Color online) Comparison of the experimental and the
  calculated low-lying energy spectra of $^{58}$Ni. The experimental data
  are taken from Refs.
  \cite{NNDC,allmond14}, while the calculated spectrum is obtained
  with the PC-PK1 force. 
  The $E2$ transition strengths are given in units of $e^2$fm$^4$.}
\label{Ni58spectra}
\end{figure}

\begin{figure}
  \includegraphics[width=8cm]{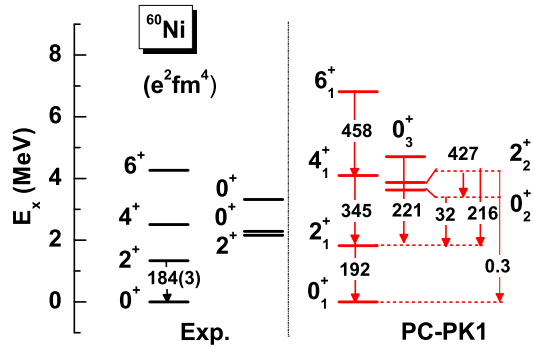}
\caption{(Color online) Same as Fig.~\ref{Ni58spectra}, but for $^{60}$Ni.}
\label{Ni60spectra}
\end{figure}

In order to facilitate the discussion on the properties of these
states, 
we collect them and make a comparison with the experimental
spectrum in Fig.~\ref{Ni58spectra}.
The results for the $^{60}$Ni are also shown in Fig. 
~\ref{Ni60spectra}.
One can see that 
the main feature of the energy spectrum and the $E2$ transition
strength from $2^+_1$ to $0^+_1$ are reproduced rather well.
These results are qualitatively similar to the results
of the previous MR-DFT calculations
with the non-relativistic Skyrme SLy4 interaction \cite{Yao15}, 
but quantitatively the present calculations with the relativistic DFT
reproduce 
the experimental data slightly better. For instance, the $B(E2)$ value
from the $2^+_1$ to the $0^+_1$ states in $^{58}$Ni was
248 $e^2$ fm$^4$ in the previous calculation\cite{Yao15}, while it is 150
$e^2$ fm$^4$
in the present calculation, that is much closer to the experimental
value of 126(8) $e^2$ fm$^4$ \cite{NNDC,allmond14}.

It is interesting to notice that 
the overall pattern of $B(E2)$ values is quite different
from what would be expected for a harmonic vibrator, even though 
the excitation energies of the 4$^+_1$, 2$^+_2$, and 0$^+_2$ states
are about twice the energy of the $2^+_1$ state. 
In particular,
the $E2$ transition from the $0^+_2$ to the $2^+_1$ states 
is much smaller than that
from the $4^+_1$ and the $2^+_2$ states to the $2^+_1$ state, 
which is similar to Cd
isotopes~\cite{Garrett08a,Garrett10a}. 
Instead, the $0^+_2$ state has a strong transition from the 
$2^+_2$ state, which clearly indicates that the $0^+_2$ state 
is not a member of the two-phonon triplet. 
Notice that the collective wave function for each of 
the $4^+_1$ and $2^+_2$ states have a similar structure to one another 
(see Fig.~\ref{Ni58PES}(b)). 
On the other hand, 
the $0^+_2$ state has a considerably different wave
function from those states, 
being dominated by the mean-field  configurations
around $\beta=0.3$. 

Compared to the $0^+_2$ state, the $E2$
transition strength from the $0^+_3$ to the $2^+_1$ states is 
much larger and is comparable to that from the $4^+_1$ and the $2^+_2$ 
states to the $2^+_1$ state. 
This fact makes 
the $0^+_3$ state a better candidate for a member of the two-phonon 
triplets, even 
though the excitation energy is a little bit large and its
collective wave function is much different from that for 
the $4^+_1$ and the $2^+_2$ states. 
A similar conclusion has been reached also with the non-relativistic 
DFT \cite{Yao15}. 

Notice that, in the harmonic oscillator limit, the $B(E2)$ value from any of 
the two-phonon triplet states to the 2$^+_1$ state is exactly twice the $B(E2)$ 
value from the 2$^+_1$ state to the ground state \cite{Greiner}. 
The calculated $B(E2)$ values shown in Figs. ~\ref{Ni58spectra} and 
~\ref{Ni60spectra}, together with the strong transition from the 2$^+_2$ to 
the 0$^+_2$ states, 
indicate a presence of large anharmonicity in the quadrupole vibrations 
in $^{58}$Ni and $^{60}$Ni. 
That is, the calculated $B(E2)$ values 
are significantly quenched from the values in the harmonic limit. 
This fact also implies that the present approach for fusion reactions 
with MR-CDFT provides a natural 
truncation scheme in the coupled-channels calculations, whereas the 
truncation of the phonon spectrum has to be introduced in an ad-hoc way 
if one employs the harmonic oscillator couplings. 

Another clear indication of anharmonicity is a finite value of 
quadrupole moment of the first 2$^+$ state. 
Experimentally, the spectroscopic quadrupole moment of the first 2$^+$ state 
has been measured to be 
$Q(2_1^+)=-10\pm 6$~$e$fm$^2$ for $^{58}$Ni and 
$Q(2_1^+)=+3\pm 5$~$e$fm$^2$ for $^{60}$Ni \cite{Lesser74}. 
The present MR-CDFT calculations yield 
$Q(2_1^+)$=+7.96 and +10.4 ~$e$fm$^2$ for $^{58}$Ni and 
$^{60}$Ni, respectively. 
Even though the sign of quadrupole moment is opposite 
to the experimental data for $^{58}$Ni, the 
MR-CDFT calculations predict a similar absolute value of quadrupole moment 
to the experimental value both for $^{58}$Ni and $^{60}$Ni. 
Notice that the average deformation for the 2$^+_1$ state is very small 
for both of these nuclei 
due to a large cancellation between the prolate and the oblate components, 
as shown in Fig. \ref{Ni58PES}. A more careful numerical treatment of 
the calculations would therefore be necessary  
in order to reproduce the correct sign of the quadrupole moment, although 
it is beyond the scope of the present paper. 

\section{Fusion of N\lowercase{i} isotopes}

In the previous section, we have seen that both $^{58}$Ni and 
$^{60}$Ni do not show a typical behavior of harmonic oscillator. 
Let us now investigate how 
the deviation of the spectrum from the harmonic limit 
affects the subbarrier fusion reactions 
of Ni isotopes. 

\begin{table*}[hbt]
  \caption{The Coulomb coupling strengths
    for the quadrupole transitions in $^{58}$Ni estimated
with the microscopic MR-CDFT
  calculations with PC-PK1 force (see Eq. (\ref{beta-gcm})). The values in the parenthesis
  are the corresponding nuclear coupling strengths obtained with Eq. (\ref{mn/mp}).
  The radius parameter of $r_0$=1.06 fm is used, and the calculated values are scaled with
  the Coulomb coupling strength for the transition from the ground state to the first 2$^+$ state,
  which is estimated with the measured $B(E2)$ value.
}
\begin{center}
\label{table:58Ni}
\begin{tabular}{c|cccccc}
\hline
\hline
& \multicolumn{6}{c}{$I'$} \\
\cline{2-7}
$I$ & 0$_1^+$ & 2$_1^+$ & 0$_2^+$ & 2$_2^+$ & 4$_1^+$ & 0$_3^+$ \\
\hline
0$_1^+$ & 0  & 0.223   & 0  & 0.0390   & 0   & 0    \\
        &        & (0.245) &        & (0.0228) &        &        \\
2$_1^+$ & 0.223  & $-$0.0457   & 0.0736  & $-$0.147   & 0.215   & 0.117   \\
        & (0.245)  & ($-$0.0311)  & (0.0668)  &($-$0.155)    & (0.229)   & (0.108)    \\
0$_2^+$ & 0  & 0.0736  & 0  & $-$0.300   & 0   & 0   \\
        &        & (0.0668)  &      &($-$0.278)    &       &         \\
2$_2^+$ & 0.0390  & $-$0.147  & $-$0.300  & 0.0873   & $-$0.0617   & 0.165   \\
        & (0.0228)& ($-$0.155)  & ($-$0.278) &(0.0777) & ($-$0.0610)  & (0.170)        \\
4$_1^+$ & 0  & 0.215  & 0  & $-$0.0617   & 0.0279   & 0   \\
        &        & (0.229)  &      &($-$0.0610) & (0.0405)  &        \\
0$_3^+$ & 0  & 0.117  & 0  & 0.165 & 0   & 0   \\
        &        & (0.108)  &      &(0.170) &   &        \\
\hline
\hline 
\end{tabular}
\end{center}
\end{table*}

\begin{table}[hbt]
  \caption{The strengths for the Coulomb coupling in $^{58}$Ni in
    the harmonic oscillator limit. Here, we have assumed that the third 0$^+$
    state belongs to the two-phonon triplet. 
}
\begin{center}
\label{table:58Ni-HO}
\begin{tabular}{c|cccccc}
\hline
\hline
& \multicolumn{6}{c}{$I'$} \\
\cline{2-7}
$I$ & 0$_1^+$ & 2$_1^+$ & 0$_2^+$ & 2$_2^+$ & 4$_1^+$ & 0$_3^+$ \\
\hline
0$_1^+$ & 0  & 0.223   & 0  & 0   & 0   & 0    \\
2$_1^+$ & 0.223  & 0   & 0  & $-$0.169   & 0.226   & 0.141   \\
0$_2^+$ & 0  & 0  & 0  & 0   & 0   & 0   \\
2$_2^+$ & 0  & $-$0.169  & 0  & 0   & 0   & 0   \\
4$_1^+$ & 0  & 0.226  & 0  & 0   & 0   & 0   \\
0$_3^+$ & 0  & 0.141  & 0  & 0 & 0   & 0   \\
\hline
\hline
\end{tabular}
\end{center}
\end{table}

We first consider the fusion reaction of two $^{58}$Ni nuclei. 
Since we assume the axial and reflection 
symmetries in the present MR-CDFT calculations, 
at this moment we are unable to describe 
both the octupole vibration, 3$^-$, and the 3$^+$ state in the three-phonon 
multiplets. We therefore consider only the quadrupole 
two-phonon excitations in each $^{58}$Ni nucleus. 
Table I summarizes the coupling strengths, 
obtained with the MR-CDFT calculation discussed 
in the previous section (see Eq. (\ref{beta-gcm})). 
We use $R_T=1.06\times 58^{1/3}$ fm for the radius of $^{58}$Ni. 
The phase of each of the collective wave functions is chosen so that 
the sign of the off-diagonal components is identical to that 
in the harmonic oscillator limit.  
The Coulomb coupling strength for the transition between the ground state 
and the first 2$^+$ state is estimated to be $\beta=0.223$ using the measured 
$B(E2)$ value with the same radius parameter. 
The calculated values of the coupling strengths shown in 
Table I have been 
scaled to this value, which amounts to multiplying a factor of 0.916 to all 
the calculated coupling strengths. 
In Table I, the values in the parenthesis are the 
nuclear coupling strengths, calculated with the theoretical value 
for $M_n/M_p$ based on Eq. (\ref{mn/mp}). 
For a comparison, we also show in Table II the Coulomb coupling strengths 
in the harmonic oscillator limit, assuming the third 0$^+$ state to be a 
member of the two-phonon triplet. 
As we have mentioned in the previous section, 
the coupling strengths in $^{58}$Ni shown in Table I reveal 
some similarity to the harmonic oscillator. 
That is, the coupling strengths from the one phonon state (the 
2$_1^+$ state)
to the two-phonon states (that is, 
the 2$^+_2$, 4$^+_1$, and 0$^+_3$ states) are 
close to those in the harmonic limit. However, there are also 
pronounced deviations from the harmonic limit as well. 
That is, 
the strong couplings are present between the 2$^+_2$ and the 0$_2^+$ 
states, and also 
between the 0$^+_3$ and the 2$^+_2$ states. 
The latter is zero in the harmonic limit, and so is the former 
unless the 0$^+_2$ is a member of the three-phonon multiplets. 

\begin{figure}[htb]
\includegraphics[scale=0.5,clip]{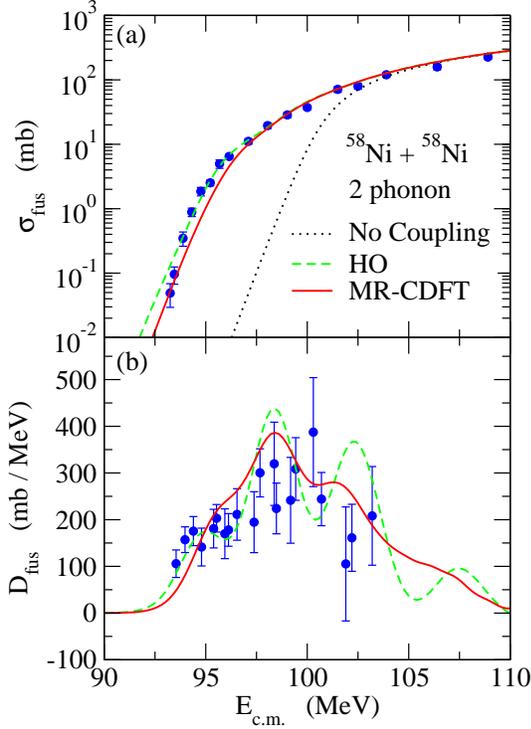}
\caption{(Color online)
The fusion cross sections (the upper panel) and 
the fusion barrier distributions (the lower panel) 
for the $^{58}$Ni+$^{58}$Ni system. 
Here, the fusion barrier distribution is 
defined as $D_{\rm fus}(E)=d^2(E\sigma_{\rm fus})/dE^2$.
The dashed line is the result of the coupled-channels calculations 
including the double quadrupole phonon excitations in each $^{58}$Ni 
nucleus in the harmonic oscillator limit, while the solid line is 
obtained by including 
the 0$^+_1$, 2$^+_1$, 0$^+_3$, 2$^+_2$, and 4$^+_1$ states with the 
coupling strengths shown in Table I. The dotted line denotes 
the result in the absence of the channel couplings. 
The experimental data are taken from Ref. \cite{Beckerman81} for 
the fusion cross sections and from Ref. \cite{Stefanini95} 
for the fusion barrier distribution. 
}
\label{fig:58ni58ni}
\end{figure}

Figures \ref{fig:58ni58ni} (a) and \ref{fig:58ni58ni} (b) show the fusion 
cross section $\sigma_{\rm fus}(E)$ 
and the fusion barrier distribution $D_{\rm fus}(E)$ 
defined as 
\cite{DHRS98,RSS91}
\begin{equation}
D_{\rm fus}(E)=\frac{d^2(E\sigma_{\rm fus})}{dE^2},
\end{equation}
for the $^{58}$Ni+$^{58}$Ni reaction, respectively. 
We use the Woods-Saxon potential with $V_0$=170.2 MeV, $r_0=1.0$ fm, 
and $a$=0.9 fm, where the radius $R_0$ is given as $R_0=r_0(A_T^{1/3}+A_T^{1/3})$. 
The fusion barrier distributions are obtained with the point difference
formula with the energy step of $\Delta E_{\rm c.m.}$ = 2 MeV in
order to be consistent with the experimental barrier distribution
extracted in Ref. \cite{Stefanini95}. 
The dashed line shows the result of the 
coupled-channels calculations including up to the double phonon states  
in the harmonic oscillator limit. 
All the mutual excitations between the projectile and the target nuclei
are included. 
On the other hand, the solid line in the figure is obtained 
with the coupling strengths shown in Table I. To this end, we include 
the 0$^+_1$, 2$^+_1$, 0$^+_3$, 2$^+_2$, and 4$^+_1$ states in the 
coupled-channels calculations.
Again, all the mutual excitation channels are taken into account. 
We use the experimental excitation energies 
for these states, that is, $\epsilon=$0, 1.454, 3.531, 2.775, and 2.459 MeV, 
respectively. 
For a comparison, the figure also shows the result of no-coupling limit by 
the dotted line.
One can see that the calculations in the harmonic limit overestimate
fusion cross sections at the two lowest energies, while the MR-CDFT
calculations underpredict fusion cross sections around 95 MeV.
On the other hand, for the energy dependence of fusion cross
sections, shown in terms of fusion barrier distribution 
in the lower panel of the figure, the MR-CDFT calculation leads to a minor
improvement by considerably smearing each peak.

\begin{figure}[htb]
\includegraphics[scale=0.5,clip]{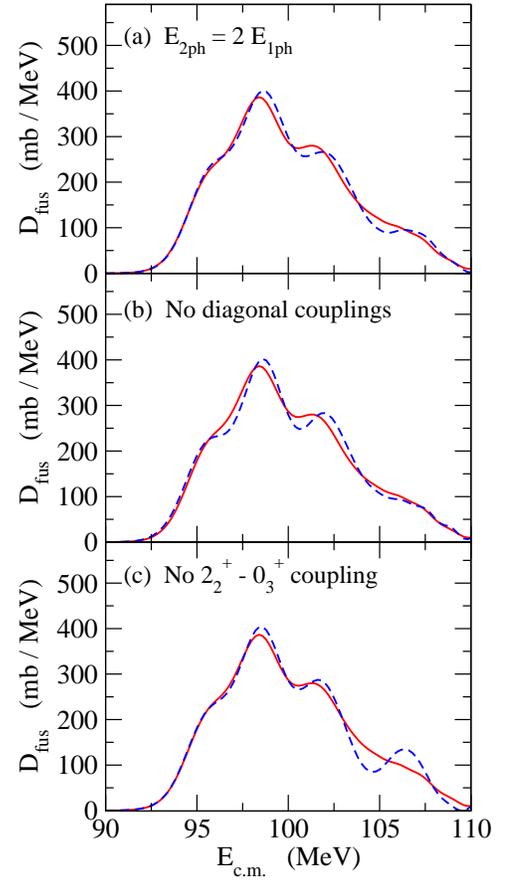}
\caption{(Color online)
  The fusion barrier distributions for the $^{58}$Ni+$^{58}$Ni reaction
  obtained with several coupling schemes. In all the panels, the solid
  line denotes the result of the MR-CDFT calculation shown in Fig. 
  \ref{fig:58ni58ni}. In the top panel, the dashed line is obtained by
  setting the energy of the two-phonon triplet to be exactly twice the
  energy of the first 2$^+$ state. In the middle panel, 
  the dashed line is obtained by setting all the diagonal couplings
  to be zero, while in the bottom panel 
it is obtained by setting the coupling strength 
to be zero 
for the transition
  from the second 2$^+$ state to the third 0$^+$ state. 
  }
\label{fig:58ni58ni-check}
\end{figure}

In order to understand the origins for the smearing in fusion
barrier distribution, we repeat the same calculations
with three different coupling schemes.
In the first scheme,
the energy of the two-phonon triplets is set to be exactly twice the
energy of the first 2$^+$ state. The result for this is shown in
Fig. \ref{fig:58ni58ni-check} (a). 
In the second scheme, all the diagonal couplings are set to be zero, while
in the third scheme the coupling strength is set to be zero between 
the 2$_2^+$ and the 0$_3^+$ states. 
The results of these schemes are plotted in Figs. 
\ref{fig:58ni58ni-check} (b) and \ref{fig:58ni58ni-check} (c), respectively. 
One can see that all of these coupling schemes lead to a more structured
barrier distribution than the full MR-CDFT calculations, and thus
all of these three effects, together with the quenching of the coupling
between the one-phonon and the two-phonon states, 
coherently contribute to the smearing in the barrier distribution.
Among them, the effect 
shown in Fig. \ref{fig:58ni58ni-check} (c)   
seems to yield the
largest effect.

\begin{figure}[htb]
\includegraphics[scale=0.5,clip]{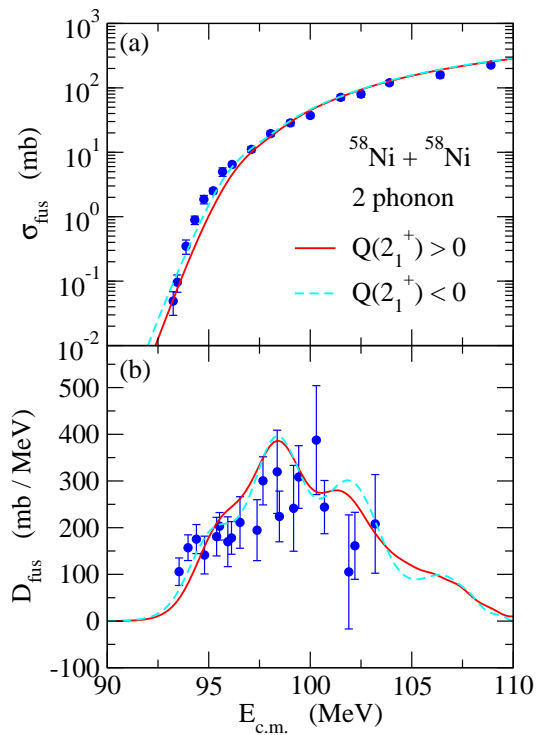}
\caption{(Color online)
  The fusion cross sections (the upper panel) and the fusion
  barrier distributions (the lower panel) for the $^{58}$Ni+$^{58}$Ni
  system obtained with the MR-CDFT method. The solid line is the same
  as that in Fig. \ref{fig:58ni58ni}, while the dashed line is obtained
  by inverting the sign of the quadrupole moment of the first 2$^+$
  state.
  The experimental data are taken from Refs. \cite{Beckerman81,Stefanini95}.
}
\label{fig:58ni58ni-q2}
\end{figure}

In connection to Fig. \ref{fig:58ni58ni-check} (b),
Fig. \ref{fig:58ni58ni-q2} shows the sensitivity of the
fusion cross section and the fusion barrier distribution to the sign of
quadrupole moment of the first 2$^+$ state.
The solid line is the result with the coupling strengths shown in Table I,
while the dashed line is obtained by changing the sign of the
quadrupole moment of the first 2$^+$ state. One can see that 
the effect of the sign of the quadrupole moment is not large, but is
certainly not negligible.
Therefore, the conclusion in Refs. \cite{HTK97,HKT98} remains the same,
that is, the sign of the quadrupole moment of an excited state
can be determined with heavy-ion subbarrier fusion reactions, when high
precision experimental data are available.
For the $^{58}$Ni+$^{58}$Ni system shown in Fig. \ref{fig:58ni58ni-q2}, 
the experimental data are 
reproduced slightly better with a negative
value of quadrupole moment of the first 2$^+$ state, which is consistent
with the experimental observation \cite{Lesser74}.

\begin{figure}[htb]
\includegraphics[scale=0.5,clip]{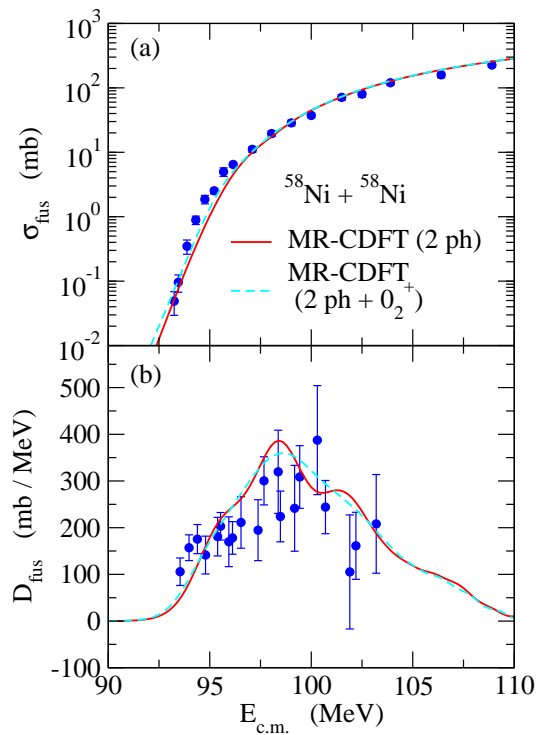}
\caption{(Color online)
  A comparison of 
  the coupled-channels calculations for the $^{58}$Ni+$^{58}$Ni
  system   with (the dashed line)
  and without (the solid line) the couplings to the second 0$^+$ state. 
  The upper and the lower panels show 
  the fusion cross sections and the fusion
  barrier distributions, respectively.
  The experimental data are taken from Refs. \cite{Beckerman81,Stefanini95}.
}
\label{fig:58ni58ni-2}
\end{figure}

Finally let us discuss the effect of the second 0$^+$ state, which couples
strongly to the second 2$^+$ state (see Table I). 
The dashed line in Fig. \ref{fig:58ni58ni-2} is obtained
by including the second 0$^+$ state in the coupled-channels
calculations in addition to the two-phonon excitations. 
On the other hand side, the solid line shows the result of the two-phonon
excitations, which is the same as that in Figs. \ref{fig:58ni58ni},
\ref{fig:58ni58ni-check} and \ref{fig:58ni58ni-2}.
Despite the strong coupling between the second 0$^+$ and the second 2$^+$ 
states, 
one can see that the main feature of the barrier distribution remains the
same even when the second 0$^+$ state is included, 
although
the 
peak structure is further smeared by the second 0$^+$ state.
This is probably because the
second 0$^+$ state is not directly coupled to the ground state. 

\begin{table*}[hbt]
  \caption{Same as Table \ref{table:58Ni}, but for $^{60}$Ni.
}
\begin{center}
\label{table:60Ni}
\begin{tabular}{c|cccccc}
\hline
\hline
& \multicolumn{6}{c}{$I'$} \\
\cline{2-7}
$I$ & 0$_1^+$ & 2$_1^+$ & 0$_2^+$ & 2$_2^+$ & 4$_1^+$ & 0$_3^+$ \\
\hline
0$_1^+$ & 0  & 0.261   & 0  & $-$0.0101   & 0   & 0    \\
        &        & (0.288) &        & ($-$0.0170) &        &        \\
2$_1^+$ & 0.261  & $-$0.0616   & 0.0478  & $-$0.148   & 0.251   & 0.126   \\
        & (0.288)  & ($-$0.0562)  & (0.0390)  &($-$0.154)    & (0.267)   & (0.117)    \\
0$_2^+$ & 0  & 0.0478  & 0  & $-$0.390   & 0   & 0   \\
        &        & (0.0390)  &      &($-$0.367)    &       &         \\
2$_2^+$ & $-$0.0101  & $-$0.148  & $-$0.390  & 0.164   & $-$0.0264   & 0.107   \\
        & ($-$0.0170)& ($-$0.154)  & ($-$0.367) &(0.151) & ($-$0.0210)  & (0.114)        \\
4$_1^+$ & 0  & 0.251  & 0  & $-$0.0264   & $-$0.0701   & 0   \\
        &        & (0.267)  &      &($-$0.0210) & ($-$0.0681)  &        \\
0$_3^+$ & 0  & 0.126  & 0  & 0.107 & 0   & 0   \\
        &        & (0.117)  &      &(0.114) &   &        \\
\hline
\hline
\end{tabular}
\end{center}
\end{table*}

\begin{figure}[hbt]
\includegraphics[scale=0.5,clip]{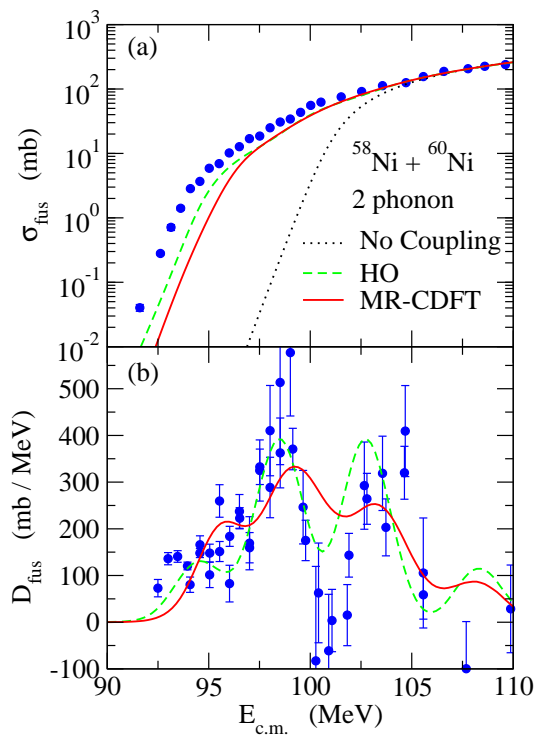}
\caption{(Color online)
Same as Fig. \ref{fig:58ni58ni}, but for
the $^{58}$Ni+$^{60}$Ni system.
The experimental data are taken from Ref. \cite{Stefanini95}. 
}
\label{fig:58ni60ni}
\end{figure}

Let us next consider the $^{58}$Ni+$^{60}$Ni reaction, for which high precision
data for fusion cross sections have been measured by
Stefanini {\it et al.} \cite{Stefanini95}.
Table III summarizes the coupling strengths for $^{60}$Ni 
obtained with the MR-CDFT calculations. The main feature of the
coupling strengths is similar to that for $^{58}$Ni, although
the collectivity is somewhat larger in $^{60}$Ni than in $^{58}$Ni.
In particular, 
strong couplings between the 0$_2^+$ and 2$_2^+$ states 
and between the 0$_3^+$ and 2$_2^+$ states 
remain qualitatively the same.
In addition, the reorientation term (that is, the self-coupling term) 
for the $2_2^+$ 
is much larger in $^{60}$Ni as compared to that in $^{58}$Ni. 
Fig. \ref{fig:58ni60ni} shows the results of the coupled-channels
calculations with the Woods-Saxon potential with $V_0$=154.5 MeV,
$r_0$=1.0 fm, and $a$=0.9 fm. Since we would like to compare
between the harmonic limit and the MR-CDFT calculations, we
consider only the two-phonon couplings, excluding the couplings to the
second 0$^+$ state.
Unlike the $^{58}$Ni+$^{58}$Ni system shown in Fig. \ref{fig:58ni58ni}, 
the fusion cross sections are
largely underestimated at energies below the Coulomb barrier by this 
calculation. 
This is probably due to the elastic two-neutron transfer process, which
is not taken into account in the present coupled-channels calculations,
as has been discussed in Ref. \cite{Stefanini95}.
Despite this, however, one may expect that the shape of fusion barrier
distribution is not much affected by transfer channels, unless a
multi-nucleon transfer process takes place (which is unlikely in the
$^{58}$Ni+$^{60}$Ni system). The effect of anharmonicity on the shape of
fusion barrier distribution is qualitatively the same as that in the
$^{58}$Ni+$^{58}$Ni system. That is, the anharmonicity largely
smears the peak structure in the barrier distribution.
Even though the agreement with the experimental barrier distribution
gets worse by including the anharmonicity effects, it is interesting
to notice that the MR-CDFT calculations appear to reproduce the more
recent (preliminary) 
data of the barrier distribution for the same
system \cite{Rodriguez10}, in a more consistent way than the 
result of the harmonic approximation. 

\begin{figure}[hbt]
\includegraphics[scale=0.5,clip]{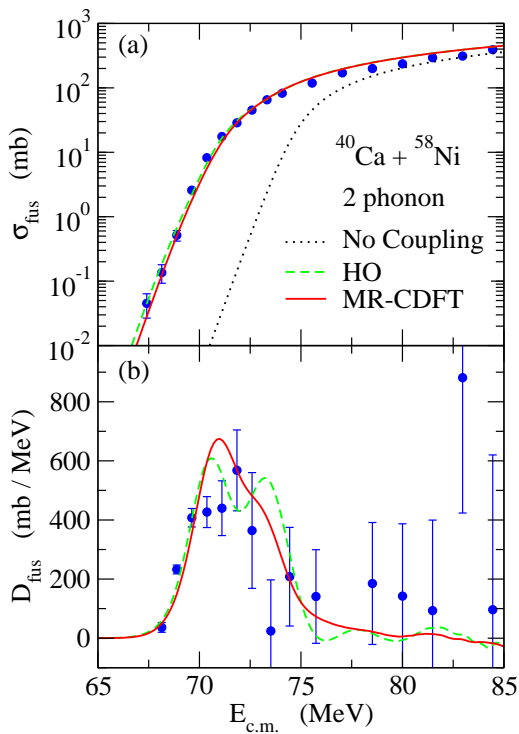}
\caption{(Color online)
Same as Fig. \ref{fig:58ni58ni}, but for
the $^{40}$Ca+$^{58}$Ni system.
The experimental data are taken from Ref. \cite{Bourgin14}. 
}
\label{fig:40ca58ni}
\end{figure}

Lastly, we briefly discuss the $^{40}$Ca + $^{58}$Ni fusion reaction. 
Fig. \ref{fig:40ca58ni} shows the results of the coupled-channels
calculations with the Woods-Saxon potential
with $V_0$=135 MeV, $r_0$=1.0 fm, and $a$=0.9 fm.
The excitations up to the two-phonon states are taken into account in the target
nucleus $^{58}$Ni while the one octupole phonon excitation is
included for the projectile nucleus, $^{40}$Ca, in the harmonic
limit. All the mutual excitations are included in the coupled-channels
calculation.
Since the charge product is small for this system, the
inclusion of the second 0$^+$ state in $^{58}$Ni leads to only a marginal
change both in the fusion cross sections and in the fusion barrier
distribution.
Both the harmonic limit and MR-CDFT calculations reproduce well the
experimental fusion cross sections \cite{Bourgin14}. 
However, in the lower panel of the figure 
one can again see that the anharmonicity effect in $^{58}$Ni smears
the fusion barrier distribution, leading to a better agreement with the
experimental fusion barrier distribution
as compared to the results in the harmonic
oscillator limit.

\section{Summary}

We have proposed the semi-microscopic approach to heavy-ion sub-barrier 
fusion reactions. 
The basic idea of this approach is to combine a multi-reference 
density functional theory (MR-DFT) to a coupled-channels calculation. 
The MR-DFT provides transition strengths 
among collective states without resorting to the harmonic oscillator model 
or the rigid rotor model. It also provides the relative sign for the 
transition matrix elements. These quantities are usually not known 
experimentally, and the MR-DFT 
provides important inputs for the coupled-channels calculations. 
The excitation energies, on the other hand, are often known well, and 
one can simply use the experimental values for them, although one could use 
the results of a MR-DFT calculation if the excitation energies are not 
known experimentally. In this paper, instead of carrying out a fully 
microscopic calculation for fusion with a double folding potential, we 
employ the semi-microscopic approach with a phenomenological Woods-Saxon 
potential, since it has been known that the double folding procedure 
does not work for subbarrier fusion reactions. 
The advantages of this approach include i) deviations from the harmonic 
limit as well as the rigid rotor limit can be taken into account, 
ii) it can therefore be applied also to transitional nuclei, which 
show neither the vibrational nor the rotatinal characters, 
iii) the sign 
of the matrix elements can be determined, iv) the nuclear coupling strengths 
can be estimated using the neutron-to-proton ratio for a transition, 
and v) a natural truncation is introduced in the coupling schemes. 

We have applied this approach to 
the $^{58}$Ni+$^{58}$Ni, $^{58}$Ni+$^{60}$Ni, and $^{40}$Ca+$^{58}$Ni fusion 
reactions. 
We have first discussed the spectrum of $^{58}$Ni and $^{60}$Ni using 
the multi-reference covariant density functional theory. We have 
found that there are both similarities and differences between 
the calculated spectra and those in the harmonic limit, even though those nuclei 
have been considered to be typical vibrational nuclei in the literatures. 
We have then discussed the effect of the anharmonicities on the fusion 
cross sections and the fusion barrier distributions. 
We have found that the anharmonicities smear the fusion barrier distributions, 
somewhat improving the agreement with the experimental data. 

In this paper, for simplicity 
we have assumed the axial and reflection symmetries 
for the shape of $^{58}$Ni and $^{60}$Ni. This has prevented us from 
including the octupole and the three quadruple phonon excitations in 
the coupled-channels calculations. 
Although the octupole excitations would not affect much the fusion 
cross sections since the excitation energy is large \cite{HT12,THAB94}, 
the three quadrupole phonon excitations may perturb 
the results presented in this paper. 
It would be an interesting future problem, 
even though it may be numerically demanding, to relieve the restriction  
for the symmetries and repeat the calculations in order to investigate 
the role of the three-phonon excitations. 

It would also be an interesting problem to extend the treatment presented 
in this paper to heavy-ion elastic and inelastic scattering. In contrast to 
fusion reactions, the double-folding approach is applicable to these 
reactions. 
One can therefore develop a fully microscopic approach to 
heavy-ion scattering using the multi-reference density functional theory. 
That approach would also be useful in applications 
to nuclear data, such as those related to the problem of 
nuclear transmutation \cite{Sukhovitskii00}. 

\section*{acknowledgements}
We thank D.J. Hinde, E. Williams, and A.M. Stefanini 
for useful discussions on the experimental data. 
This work was partially supported by the National Natural Science Foundation of China under 
Grant Nos. 11305134,  11105111, and the Fundamental Research Funds for the Central 
University (XDJK2013C028).

\end{document}